\documentclass[twocolumn,showpacs,preprintnumbers,amsmath,amssymb,floatfix]{revtex4}
\usepackage{graphicx}
\usepackage{dcolumn}
\usepackage{bm}
\begin{document}
\def\x{{\mbox{\boldmath$x$}}}
\def\u{{\mbox{\boldmath$u$}}}
\def\r{{\mbox{\boldmath$r$}}}
\def\z{{\mbox{\boldmath$z$}}}
\def\y{{\mbox{\boldmath$y$}}}
\def\k{{\mbox{\boldmath$k$}}}
\def\L{{\mbox{\boldmath$L$}}}
\def\unitr{{\mbox{\boldmath$\hat r$}}}
\def\t{\vartheta}
\def\f{\varphi}
\def\nab{{\bf \nabla}}
\def\eps{{\epsilon}}
\def\ba{\begin{eqnarray}}
\def\ea{\end{eqnarray}}
\def\be{\begin{equation}}
\def\ee{\end{equation}}
\title{Response maxima in modulated turbulence}
\author{Anna von der Heydt$^{1,2}$, Siegfried Grossmann$^1$, and 
Detlef Lohse$^2$}
\affiliation{
$^1$Fachbereich Physik, Philipps-Universit\"at Marburg, Renthof 6,
35032  Marburg, Germany\\
$^2$Department of Applied Physics  and J.\ M.\ Burgers Centre for
Fluid Dynamics, 
University of Twente, 7500 AE Enschede,
The Netherlands}
\date{\today}
\begin{abstract}
{Isotropic and homogeneous turbulence driven by an energy input  
modulated in time is studied within a variable range
mean-field theory. The
response of the system, observed in the second order 
moment of the large-scale velocity difference
$D(L,t)=\langle\!\langle(\u(x+L)-\u(x))^2\rangle\!\rangle\propto
Re^2(t)$, is calculated for varying modulation frequencies $\omega$ and
weak modulation amplitudes. For low
frequencies the system follows the modulation of the driving with
almost constant amplitude, whereas for higher driving frequencies the
amplitude of the response decreases on average $\propto 1/\omega$. In addition,
at certain frequencies the amplitude of
the response either almost
vanishes or is strongly enhanced. These frequencies are connected with
the frequency scale of the energy cascade and multiples thereof. }
\end{abstract}
\maketitle
\section{Introduction}
Many turbulent flows are characterized by time dependent forcing.
E.g. the  atmosphere of the earth is driven by
the heating through the radiation from the sun, the blood flow in the
arteries by the heart beats, etc.  Also
technical flows  like the flow in the intake of a combustion engine
are periodically forced. Another example are estuaries and adjacent
coastal waters, where tidal straining leads to a periodic alternation of
stratification and turbulent mixing of saline and fresh water 
\cite{rip01}. This results in a periodically varying energy dissipation in
the upper water layers with a 12 hour period. 

The effect of a periodically increasing and decreasing energy input on
turbulent flow 
depends on the frequency of the driving. This has been studied in
reference \cite{sco01} for a turbulent channel flow where the modulations of
the input rate are
generated near the wall. It was found that for high frequencies these
oscillations are strongly damped with distance from the walls, such that they do not reach the
inner part of the 
logarithmic boundary layer. Another example is Rayleigh-Benard
convection: the interaction between the large scale circulating flow and
the thermal plumes detaching from the upper and the lower boundary
layers acts as a stochastically influenced time-dependent forcing on
the turbulent flow  in the inner region of the
cell, as recently shown in \cite{qiu00a,qiu01b,qiu01a}. 
In a von K\'arm\'an flow between two coaxial corotating disks 
\cite{lab96b,aum00}, the energy input rate is not constant if the
disks are kept rotating at constant speed, but is periodically varying with a
geometry-dependent frequency due to a coherent vortex precessing
around the axis of rotation. In this case it was also shown, that the
statistical properties of the turbulent fluctuations are affected by
the time dependence of the mean flow. However, the averaged velocity
power spectrum still shows Kolmogorov scaling over a broad frequency
range, in addition to a low frequency peak corresponding to the
oscillation of the mean flow. 

These results raise the question how global quantities of a turbulent
flow, like e.g. the total energy or the Reynolds number, respond to a
time dependent energy input. This problem is the subject of the
present paper. From a more fundamental point of view, studying
modulated turbulence will give more insight into the time scales in
particular of the turbulent energy cascade. 
 
In a previous study \cite{loh00}, the time evolution of the Reynolds
number in a  
periodically kicked flow was analyzed. If the kicking strength and the
kicking frequency are large enough, the Reynolds number grows and
saturates on a level, which depends on the frequency and the kicking
strength. The theoretical results from \cite{loh00} have later
been verified numerically in reference \cite{hog01}. 

In this present paper, we study a related type of forcing. Rather than
periodically kicking the boundary conditions of homogeneous, isotropic
turbulence as in \cite{loh00}, we force the flow through a
time-dependent modulation of the energy input rate $e_{in}(t)$ on the
outer length scale $L$,  
\begin{equation}
\label{eint}
e_{in}(t)=e_0(1+e\sin (\omega t)).
\end{equation}
This means that the flow is stationarily stirred ($\propto e_0$) 
to maintain the turbulent flow and, in addition, a time-dependent
modulation of the 
forcing ($\propto e_0e$) is applied, $0\leq e\leq 1$.  The response of
the  system to
the time-dependent stirring can be observed e.g. in the second order velocity
structure function of the flow field, in particular at the outer scale
$L$,  $D(L,t)=\langle\!\langle (\u(\x+\L,t)-\u(\x,t))^2
\rangle\!\rangle$.  This $D(L,t)$ is equivalent  to a Reynolds number,
which  we define as $Re=u_{1,rms}L/\nu$. Here, $u_{1,rms}(t)$
is the rms of one component of the velocity, varying with time
$t$. Then, disregarding correlations on scale $L$, 
\begin{equation}
\label{DRe}
D(L,t)=2\langle\!\langle \u^2\rangle\!\rangle=6u_{1,rms}^2(t)
=\frac{6\nu^2Re(t)^2}{L^2}.
\end{equation} 

The energy put into the
system at time $t$ will travel down the energy cascade
towards smaller 
scales and will, on average, be dissipated at time $t+\tau$, i.e.,
with a mean time delay $\tau$.  In
other words, the dissipation at time $t$ depends on how much energy
has been in the large scales  
at time $t-\tau$. We approximately describe the relevant time scale $\tau$ for
the cascade process by the large eddy turnover time $\tau_L$ at that time $t-\tau$, 
\begin{equation}
\tau\simeq\tau_L={L\over u_{1,rms}(t-\tau)}={L\over\sqrt{D(L,t-\tau)/6}}.
\label{tau} 
\end{equation}
More accurately, the time scale of the energy cascade is given by the
sum over the eddy turnover times on all decay steps,
$\tau\simeq\sum_n\tau_n$.  In this sum, the largest contribution
is the largest eddy turnover time $\tau_L$. For K41 scaling the smaller eddies
$r_n/L=\delta^n$, where $0<\delta<1$, have turnover times
$\tau_n=\tau_L \delta^{2n/3}$. Thus 
$\tau=\tau_L\sum_n \delta^{2n/3}\equiv \tau_L a$. The common choice
$\delta=1/2$ implies $a\simeq 2.7$. Putting into intermittency
corrections gives slightly smaller values of $a$. 
In this present paper we shall discuss the influence of $a$ by
comparing the limiting cases $a=2.7$ and $a=1$. Experimentally, in
principle the parameter $a$ could be measured by analyzing the
positions, heights and widths of the response maxima, thus giving
information about the energy cascade time. 

If the external modulation period $\omega^{-1}$ is much larger than
this intrinsic time scale $\tau$, $\omega\tau\ll1$, the turbulent flow
will have time to adjust and will follow the periodic variations of
the stirring. If, on
the other hand, $\omega^{-1}$ is decreased and becomes much smaller
than $\tau$, the system can follow less and less, and feels, at small scales, an
average time-independent energy input.  

We calculate the time dependence of the response $D(L,t)-D_0(L)$ to a
periodically modulated energy input rate, Eq. (\ref{eint}),
within a variable scale mean-field theory \cite{eff87} for various driving
frequencies $\omega$. Here, $D_0(L)$ is the second order structure
function  for a stationary energy input rate $e_0$. 
In general, the energy flow rate through the system is an
intermittently fluctuating quantity. Therefore, the cascade time 
as well as the response of the system are
fluctuating. These fluctuations are neglected by the mean-field theory
in the present study. However, on average these fluctuations result in
a mean downscale transport of energy which controls the overall
properties of the flow.  Therefore, we believe that within this
mean-field approach we can grasp the main
features of the flow correctly. 

The method is explained in the next section. The behavior of the
response as a function of the driving frequency $\omega$ in the case of
weak modulations of the energy input rate 
is analyzed in section \ref{response}. In section
\ref{EGleichung} we discuss an alternative way to introduce time
dependence into the system. 
The slightly different case of a modulated driving force instead of a modulated energy input rate is presented in section \ref{modf}. 
We summarize our results in section \ref{discussion}.

\section{Method and Model}
\label{method}
In reference \cite{eff87} an energy balance 
equation for the second order velocity structure function
$D(r)=\langle\!\langle(\u(\x+\r)-\u(\x))^2\rangle\!\rangle$ for 
stationary, homogeneous, and isotropic turbulence has been derived within a
variable  range mean-field
theory. Here, $\u$ is the velocity and the brackets $\langle\!\langle
... \rangle\!\rangle$
denote the ensemble average. One of the essentials of this theory is
to divide the velocity field into a (spatially averaged) {\it
superscale} velocity $\u^{(r)}$ and a (strongly
fluctuating) {\it subscale} velocity $\tilde \u^{(r)}$. The spatial average
is performed over a sphere of variable radius $r$, and will be denoted as 
$\u^{(r)}(\x)\equiv\langle \u(\x+\y)\rangle^{(r)}_y\equiv {3\over 4\pi
r^3}\int_{|\y|\leq r}d^3y\ \u(\x+\y)$.

The energy input rate $e_{in}$, 
which in the statistically stationary situation equals the total energy 
dissipation rate $\epsilon$,  
is balanced in accordance with the super- and subscale decomposition by 
the energy dissipation rate on all scales larger than  $r$ complemented by  
the energy transfer across scale $r$ from the super- to the subscales
of $r$. In a simplified version the derived energy balance equation
reads: 
\begin{equation}
\label{effgngl1}
e_{in}=\epsilon={3\over 2}\left(\nu +\frac{D(r)^2}{b^3\epsilon}\right)
{1\over r}{d\over dr}D(r),
\end{equation}
where $\nu$ is the kinematic viscosity and $b$ the Kolmogorov constant.  
In the viscous subrange (VSR), where $r$  is smaller than the
Kolmogorov length scale $\eta$, $r<\eta$, the dissipation term,
i.e., the first term on the rhs of Eq. (\ref{effgngl1}), is
dominating, and therefore the solution of
Eq. (\ref{effgngl1}) is $D(r)=\epsilon r^2/3\nu$.  
In the inertial subrange, instead, where $\eta\ll r\ll L$, most
of the energy of the eddies is transfered down-scale. This
energy transfer rate $E_t$, which is given by the second term on the rhs of
Eq. (\ref{effgngl1}), is determined by the decorrelation rate
$\tilde\Gamma(r)$ of the subscale eddies, which itself is mainly
governed by the energy dissipation rate $\epsilon$, see
\cite{eff87} for details. Note again that in the stationary case the energy
dissipation rate equals the energy input rate, $\epsilon=e_{in}$. In the ISR
the second term on the rhs is the leading
one. Then the solution of Eq. (\ref{effgngl1}) is $D(r)=b(\epsilon
r)^{2/3}$. The full energy rate balance equation (\ref{effgngl1})
interpolates between these two limits. The Kolmogorov constant $b$ can be
calculated within this theory to be $b=6.3$
which is consistent with the experimental value \cite{my75,sre95,sre97,pop00}. 

In our case the flow is not stationary but experiences a modulated energy 
input rate $e_{in}(t)$. Therefore, $e_{in}$, the structure
function $D(r)$, and the dissipation rate $\epsilon$ in Eq. 
(\ref{effgngl1}) will depend on time. Furthermore, an additional term
on the rhs of Eq. (\ref{effgngl1}) appears, taking into account
the non-stationarity of the flow:
\begin{eqnarray}
\label{effgngl2}
e_{in}(t)&=&{3\over 2}\left(\nu +
\frac{D(r,t)^2}{b^3\epsilon(t)}\right)
{1\over r}{\partial\over \partial r}D(r,t)
\\\nonumber
&&+{1\over
2}{\partial\over\partial t}\langle\!\langle 
\u^{(r)}(\x,t)\cdot\u^{(r)}(\x,t)\rangle\!\rangle.
\end{eqnarray}
The correlation of the superscale velocities can be written as
$\langle\!\langle\u^{(r)}(\x,t)\cdot\u^{(r)}(\x,t)\rangle\!\rangle=
\langle\!\langle\u^2(\x ,t) \rangle\!\rangle-{1\over 2}\langle\langle
D(y_1+y_2,t)\rangle^{(r)}_{y_1}\rangle^{(r)}_{y_2}$. Following the
arguments in \cite{eff87} for the derivation of Eq. (\ref{effgngl1}),
we neglect multiple spatial averaging, i.e., $
\langle\langle D(y_1+y_2,t)\rangle^{(r)}_{y_1}\rangle^{(r)}_{y_2}\simeq
\langle D(y,t)\rangle^{(r)}_{y}$. 

In the stationary case the energy
dissipation rate $\epsilon=\nu\langle\!\langle\frac{\partial
u_i}{\partial x_j}\frac{\partial u_i}{\partial x_j}\rangle\!\rangle$
can be related to the large scale quantities by 
\begin{equation}
\label{eps1}
\epsilon=c_{\epsilon}\frac{u_{1,rms}^3}{L}=
c_{\epsilon}(D(L))\frac{D(L)^{3/2}}{6^{3/2}L}.
\end{equation}
Extending this expression to the time-dependent case, we have to take
into account that the energy which is fed into the system on large
scales at a time $t$
will be dissipated on small scales at a later time
$t+\tau$. We model this as follows: The energy dissipation rate at time $t$ 
is assumed to depend on the large scale quantities at time $t-\tau$:
\begin{equation}
\label{eps2}
\epsilon(t)=
c_{\epsilon}(D(L,t-\tau))\frac{D(L,t-\tau)^{3/2}}{6^{3/2}L}. 
\end{equation}
$c_{\epsilon}$ is a dimensionless function which is approximately
constant ($\simeq 1$) for very large Reynolds
numbers \cite{sre84,sre98}. 
In \cite{loh94a,gro95,sre95d} it was shown that in general
$c_{\epsilon}$  depends on
the Reynolds number, and therefore on $D(L)$. We here use
an approximation  of the expression
derived in \cite{loh94a} for high Reynolds numbers:
\begin{eqnarray}
\label{ceps}
c_{\epsilon}(D(L))&=&{9\over Re}+\sqrt{\left({6\over
b}\right)^{3}+\left({9\over  Re}\right)^{2}}\\\nonumber
&\simeq& \left({6\over b}\right)^{3/2}+\frac{9}{Re}=
\left({6\over b}\right)^{3/2}+9{\nu\over
L}\sqrt{6\over {D(L)}}.
\end{eqnarray}
The delay time $\tau$ is determined by the implicit time-delay
equation (\ref{tau}).
Assuming that the solution of Eq. (\ref{effgngl1}) in the ISR,
$D(r)=b(\epsilon r)^{2/3}$, is valid up to $r=L$, we can write
$D(r)=\left(\frac{r}{L}\right)^{2/3}D(L)$. Within our model, where we
connect small and large scale quantities at different times, the
structure function on scale $r<L$ at time $t$ will depend on the large
scale structure function at an earlier time $t-\tau$, i.e., we introduce
$D(r,t)=\left({r\over L}\right)^{2/3} D(L,t-\tau)$ into Eq. 
(\ref{effgngl2}). After multiplying with $r$, 
Eq.  (\ref{effgngl2}) can be integrated from $r=0$ up to the outer length
scale $r=L$: 
\begin{eqnarray}
\nonumber
{1\over 4}{d\over dt}(D(L,t)-\alpha D(L,t-\tau))&=&
-\frac{D(L,t-\tau)^{3/2}}{Lb^{3/2}} 
\\\nonumber
&&
-\frac{3\nu D(L,t-\tau)}{2L^2}
\\
&&
+e_{in}(t),
\label{inteq}
\end{eqnarray}
where $\alpha={27\over 44}$ originates from the integration.  
In \cite{eff87} it has been shown that, in the isotropic and homogenous case, 
$e_{in}$ is independent of the scale $r$ as the forcing is assumed 
to act on the largest scale $L$ only. 
In the stationary case the lhs of Eq. 
(\ref{inteq}) vanishes, and together with Eqs. (\ref{eps2}) and
(\ref{ceps}), Eq.  (\ref{inteq}) corresponds to
$\epsilon=e_{in}$. 
Eq.  (\ref{inteq}) contains only large scale
quantities. Effects of fluctuations in the energy input rate on the
statistical properties of the turbulent flow as observed in
\cite{lab96b} would influence the
scaling behavior of $D(r,t)$ on intermediate scales  $r$ and therefore lead to 
different values of the factor $\alpha$, but the structure of
Eq. (\ref{inteq}) would remain the same. 

Using Eq. (\ref{DRe}), we express the second order structure function 
$D(L,t)$ in
Eq. (\ref{inteq}) in terms of 
the Reynolds number $Re(t)$: 
\begin{eqnarray}
{L^2\over\nu}\!{d\over dt}(Re^2(t)\!-\!\alpha Re^2(t\!-\!\tau))\!&\!=\!&\!
-{2\over 3}\!\left({6\over b}\right)^{3/2}\!(Re^2(t\!-\!\tau))^{3/2}
\nonumber\\\label{regl1}
&&-6Re^2(t-\tau)
\\\nonumber
&&+{2\over 3}\frac{e_0 L^4}{\nu^3}(1+e\sin{\omega t}).
\end{eqnarray}
Here, we have inserted the time-dependent energy input
rate, Eq. (\ref{eint}).  
In the case a of constant energy input rate, i.e., $e=0$, Eq. (\ref{regl1}) 
simplifies to  
\begin{equation}
\label{re0gl}
0=-\frac{2}{3}\left({6\over b}\right)^{3/2}Re_0^3-6Re_0^2+
\frac{2}{3}\frac{L^4}{\nu^3}e_0,
\end{equation}
relating the stationary Reynolds number $Re_0$ to the stationary input 
rate, $\frac{L^4}{\nu^3}e_0(Re_0)=c_{\epsilon}(Re_0)Re_0^3$. 
Introducing the reduced Reynolds number $R(t)\equiv Re(t)/Re_0$ and the 
non-dimensional time $t/\tau_L^0$ as $t$ (analogously for $\tau$ 
and $\omega$), 
Eq. (\ref{regl1}) becomes 
\begin{eqnarray}
\frac{d(R^2(t)\!-\!\alpha R^2(t\!-\!\tau))}{dt}\! 
&\!=\!&\!-{2\over 3}\left({6\over b}\right)^{3/2}(R^2(t\!-\!\tau))
^{3/2}
\nonumber\\\label{regl2}
&-&\frac{6}{Re_0}R^2(t-\tau)
\\\nonumber
&+&\!\left(\!{2\over 3}\!(6/b)^{3/2}+\!\frac{6}{Re_0}\!\right)
\!(\!1\!+\!e\sin{\omega t}\!). 
\end{eqnarray}
Here, $\tau_L^0=\frac{L}{u^0_{1,rms}}$ is the large eddy turnover
time of the stationary flow. 
$R(t)$ is of order one. The delay time $\tau$ in units of the time
scale $\tau_L^0$ is given by 
\begin{equation}
\label{tauR}
\tau=\frac{a}{R(t-\tau)}. 
\end{equation}
Eq. (\ref{regl2}) describes the time evolution of $R^2(t)$,  
which is the square of the Reynolds number of a
flow exposed to a modulated energy input rate (Eq. (\ref{eint})), 
normalized by the square of
the  Reynolds number of a flow where only a constant, time-independent,
forcing is applied. 

\section{Response of turbulent flow to energy input rate modulations}
\label{response}

\subsection{General trend}
\label{trend}

In the present study we shall restrict ourselves to the case of weak 
amplitude modulation, i.e.,  $e$ in Eq. (\ref{eint}) is small. Then
we  expect that also the oscillating response 
\begin{equation}
\Delta(t)\equiv R^2(t)-1
\label{resp}
\end{equation}
has a small amplitude, and 
we can linearize Eq. (\ref{regl2}). The time delay $\tau$ is 
approximated by a time-independent constant which in our time units
$\tau_L^0$ is simply $a$. This
approximation is justified as long as $|\Delta|\ll1$. In section
\ref{tauapprox} we shall discuss the limits of this
approximation. We first consider $a=1$ which means that the cascade
time $\tau$ is taken as 
the large eddy turnover time $\tau_L^0$.  The
resulting equation of motion for the response $\Delta(t)$,
\begin{eqnarray}
{d\over dt}(\Delta(t)-\alpha\Delta(t-\tau))&=&
-\left(\left(\frac{6}{b}\right)^{3/2}+\frac{6}{Re_0}\right)\Delta(t-\tau)
\nonumber\\
&&+\left({2\over 3}(\left(\frac{6}{b}\right)^{3/2}+\frac{6}{Re_0}\right)e\sin{\omega t},\nonumber\\
\label{lineq}
\end{eqnarray}
can be solved analytically. 
The solution to the linear equation (\ref{lineq}) can be calculated
using the ansatz:
\begin{equation}
\label{ansatz}
\Delta(t)=eA(\omega)\sin{(\omega t+\phi)}.
\end{equation}
Here, $A(\omega)$ is the amplitude, and $\phi$ is the phase shift of
the response which also depends on $\omega$. 
Inserting this expression into Eq. (\ref{lineq}) gives the explicit
solution of the linear response equation (\ref{lineq}):
\begin{widetext}
\begin{equation}
\label{dRt}
\Delta(t)=e
\frac{\left({2\over 3}\left(\frac{6}{b}\right)^{3/2}+\frac{6}{Re_0}\right)}{\omega}
\frac{\Big[-\cos{\omega t}+\alpha\cos{\Big(\omega
(t+\tau)\Big)}+\frac{\left(\frac{6}{b}\right)^{3/2}+{6\over Re_0}}{\omega}\sin{\Big(\omega(t+\tau)\Big)}\Big]}
{(1+\alpha^2+\left(\frac{\left(\frac{6}{b}\right)^{3/2}+\frac{6}{Re_0}}{\omega}\right)^2-2\alpha\cos{\omega\tau}-2\frac{\left(\frac{6}{b}\right)^{3/2}+\frac{6}{Re_0}}{\omega}\sin{\omega\tau})}.
\end{equation}
\end{widetext}
In the following, we set the Kolmogorov constant $b=6$ 
for simplicity, which is near to the calculated value 6.3 \cite{eff87} 
and to the experimental value in the range $6-9$
\cite{my75,sre95,sre97,pop00}. To recover the expressions for
a general $b$ one has to replace in the following results the terms
$(1+{6\over Re_0})$ and $({2\over 3}+{6\over Re_0})$ by 
$((6/b)^{3/2}+{6\over Re_0})$ and
$({2\over 3}(6/b)^{3/2}+{6\over Re_0})$, respectively.   
The mean amplitude of the response is determined by the 
energy input rate $({2\over 3}+{6\over Re_0})e$, i.e., the last term on
the rhs of Eq. (\ref{lineq}). The time derivative on the lhs of
Eq. (\ref{lineq}) leads to a mean decrease of the amplitude as
$1/\omega$. Due to
the two terms in Eq. (\ref{lineq}) containing the time delay $\tau=a$,
corresponding terms in the second fraction of the solution
(\ref{dRt}) appear, $\propto\alpha$ and $\propto (1+{6\over Re_0})/\omega$,
respectively, which, by the periodic dependence on 
$\omega \tau$ induce a periodic variation of the amplitude with the
frequency $\omega$. For low frequencies the terms $\propto
(1+6/Re_0)/\omega$, originating from the first term on the rhs of
Eq. (\ref{lineq}), dominate, whereas for high frequencies the terms
$\propto\alpha$, due to the second term on  the lhs of
Eq. (\ref{lineq}), become more important. The latter, in particular,
lead to a periodic variation of the response amplitude up to very high
frequencies. 

\begin{figure}
\includegraphics[width=\columnwidth]{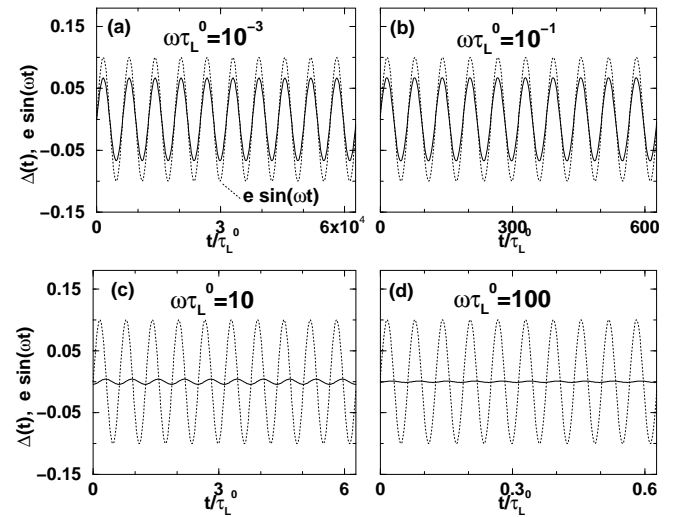}
\caption{Response $\Delta(t)$ (solid lines) for four different
modulation frequencies $\omega$, the time dependent part of the
energy input rate, $e_{in}(t)/e_0-1$ (dotted lines).
The modulation amplitude is $10\%$  of the constant 
input rate, $e=0.1$, and the Reynolds number of the stationary system
is chosen as $Re_0=10^4$.  (a) $\omega\tau_L^0=10^{-3}$, (b)
$\omega\tau_L^0=0.1$,  (c)
$\omega\tau_L^0=10$, (d) $\omega\tau_L^0=100$.}
\label{4freq-e01}
\end{figure}

The linear response $\Delta(t)\propto e$ of the flow (with
$Re_0=10^4$) is plotted in
Fig.\ref{4freq-e01} for four different modulation frequencies.  
Also the modulation of the energy input rate, 
$e_{in}(t)/e_0-1$ is plotted in Fig.\ref{4freq-e01}.
The deviation of the Reynolds number from its stationary value $Re_0$, 
$\Delta(t)=(Re^2(t)-Re_0^2)/Re_0^2$, oscillates with the same frequency
as the driving, for all frequencies $\omega$. 
The amplitude $A$ of this oscillation depends on
the frequency. For the
two small modulation frequencies, $\omega=10^{-3}$  and $\omega=10^{-1}$, 
the amplitude of the
response $\Delta(t)$ is nearly the same, about two thirds of the
amplitude $e$ of the driving.  For higher
frequencies, the amplitude $A$ of the response decreases. In the
case of $\omega=10$ we observe a phase shift  between the forcing and
the resulting response.
\begin{figure}
\includegraphics[width=\columnwidth]{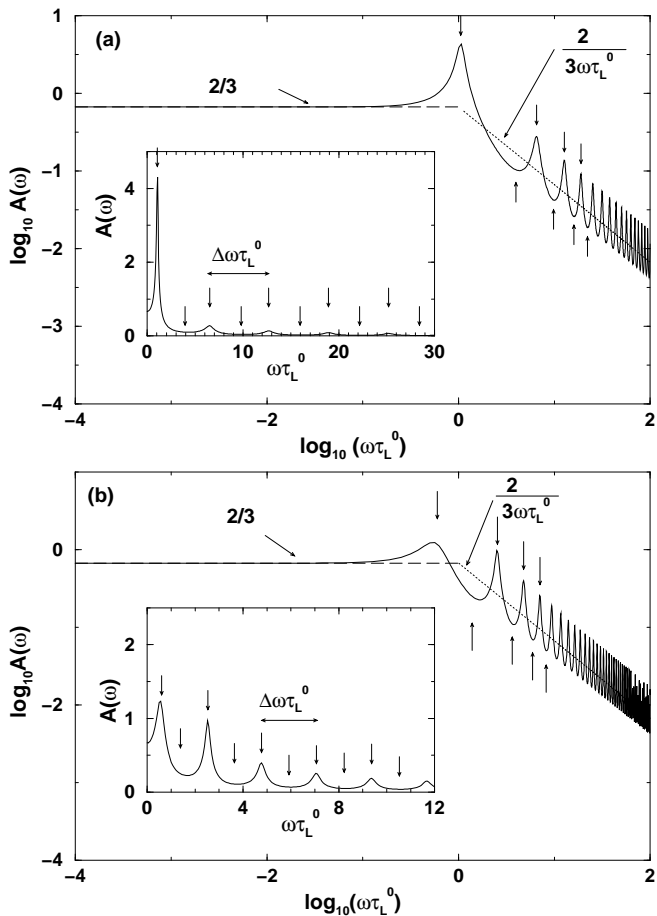}
\caption{(a) Amplitude $A$ of the response $\Delta(t)$ as a
function of 
the driving frequency $\omega$ (log-log-scale) for weak modulations
($e=0.1$) of  the 
input rate $e_{in}$, and $Re_0=10^4$. The time scale of the energy cascade is set to $\tau=a=1$. The dashed line denotes the low
frequency limit of the oscillation amplitude, $2/3$, and the
dotted line corresponds to the mean trend of the high frequency limit, 
$\frac{2}{3}\frac{1}{\omega\tau_L^0}$. Inset: linear-scale-plot of
the response amplitude versus frequency. The small arrows indicate
the frequencies $\omega_r$ (in units of $\tau_L^0$) of the response
extrema 
calculated from the extrema of
the denominator in Eq. (\ref{dRt}). The horizontal arrow denotes the
frequency distance $\Delta\omega$ (in units of $\tau_L^0$)  between two frequencies for which the
amplitude is maximal (or minimal). It is $\Delta\omega\simeq
2\pi/\tau$ for high frequencies. (b) Same as (a) but with a cascade time
scale $\tau=a=2.7$  different from the large eddy turnover time
$\tau_L^0$. Note the shift of the response maxima, the less pronounced
height and greater width of the first, and the more pronounced second
response peak.}
\label{logA}
\end{figure}
Fig.\ref{logA}a shows the amplitude $A(\omega)$ as a function of the driving
frequency for $Re_0=10^4$. For low frequencies the amplitude remains
constant, and is two thirds,
whereas for large frequencies the amplitude of
the response $\Delta(t)$ decreases $\propto 1/\omega$. 
In addition to this decrease we note certain frequencies for which
the response amplitude becomes large or very small. The distance between 
two maxima or two minima of the amplitude
is nearly constant, see the inset of Fig.\ref{logA}a. 
This periodic behavior in the $\omega$-dependence of the response
amplitude is due to the time delay
$\tau$. We shall explain this in the next section.

There are three time scales in the solution (\ref{dRt}) of Eq. 
(\ref{lineq}): The large eddy turnover time, by definition 1, the time
delay $\tau=a$, which represents the cascade time,  and the
time scale of the external modulation $1/\omega$.
If the modulation time scale is much larger than the large eddy
turnover time, $1/\omega\gg 1$, i.e., if the driving frequency
is very small, then the solution (\ref{dRt}) can be approximated by
\begin{equation}
\label{lowf}
\Delta(t)\simeq e{2\over 3}\sin{\Big(\omega(t+\tau)\Big)}.
\end{equation}
We conclude $A=2/3$, while the phase $\phi=\omega\tau$ is linear in
$\omega$ for small frequencies. 

If, on the other hand, the modulation frequency becomes very large, i.e,
the time scale of the driving is much smaller than $1$, we see
from Eq. (\ref{dRt}) that the amplitude of $\Delta$ decreases
as $\propto 1/\omega$:
\begin{equation}
\label{env}
\Delta(t)\simeq e\frac{({2\over3}+{6\over Re_0})}{\omega}
\frac{\Big[-\cos{\omega
t}+\alpha\cos{\Big(\omega(t+\tau)\Big)}\Big]}{1+\alpha^2-2\alpha\cos{\omega
\tau}}.
\end{equation}
The mean trend $\propto \frac{({2\over3}+{6\over Re_0})}{\omega}
\simeq\frac{2}{3\omega}$ of this high frequency limit is also 
plotted in Fig.\ref{logA}a. 
The crossover between the regimes of Eq. (\ref{lowf}) and
(\ref{env}) takes place at $\omega_{cross}\simeq
1$. This can be seen in Fig.\ref{logA}a. The crossover frequency  
is not changed by taking into account the cascade time $\tau=a\not=1$,  
as can be seen in Fig.\ref{logA}b which shows the response
amplitude as a function of frequency for $a=2.7$. 

We have considered here only the case, where the Kolmogorov constant
$b=6$. For a general $b$, the crossover frequency is at
$\omega_{cross}\simeq \left(6/b\right)^{3/2}$, as can be seen from the
solution (\ref{dRt}). This means, that the crossover from the regime
of constant amplitude to the regime of $1/\omega$-decay takes place at
a smaller frequency if $b$ is larger. The positions of the response
maxima, however, are only slightly shifted by a different $b$. 

In conclusion, as long as the modulation frequency of the 
energy input rate is smaller than $1$, i.e., the large eddy turnover time is shorter than the period of the forcing,
the system has time to follow  the periodic modulations with an almost
constant amplitude. 
For higher frequencies instead, the oscillations become too fast for the 
system to follow, and therefore, the response becomes weaker and
weaker, and phase shifted. Then the system experiences the fast modulation 
more and more as a constant average energy input, and the oscillations of the 
response vanish as $1/\omega$. 
This high frequency behavior has also been found for spin systems
driven by an oscillating magnetic field \cite{rao90}.

\subsection{Response maxima}
\label{resonances}

In Fig.\ref{logA} we have seen that there are certain frequencies for
which the amplitude of the response becomes large or very small. 
Mathematically, these response extrema  
originate from the minima and maxima of the denominator in Eq. 
(\ref{dRt}),
\begin{eqnarray}
N(\omega)=&\omega&\Big[1+\alpha^2+\left(\frac{1+{6\over Re_0}}{\omega}\right)^2
\\\nonumber
&&-2\alpha\cos{\omega\tau}
-2\frac{1+{6\over Re_0}}{\omega}\sin{\omega\tau}\Big].
\end{eqnarray}
We calculate the extrema of $N(\omega)$ numerically. The first few of
them are indicated by the small arrows in Fig.\ref{logA}a. The lowest
frequency is near to $\omega_{r1}\simeq\pi/3\tau\simeq 1$. There, 
the first and strongest maximum of the response can be observed, where the 
amplitude becomes as high as $A\simeq 4.2$.  Note,
that this frequency is nearly equal to the crossover frequency
$\omega_{cross}$ between the low and high frequency regimes
of Eq. (\ref{lowf}) and (\ref{env}) only in this particular case, where
$a=1$.  If we assume an energy cascade time $\tau=a=2.7$ the
frequencies of the  maxima are shifted towards smaller
frequencies. The height of the first maximum is decreased, i.e., 
$A\simeq1.2$, whereas the
height of the following maxima is slightly increased, see Fig.\ref{logA}b.
For very
large frequencies, $\omega\gg1$, we can estimate the
frequencies of the response extrema also analytically. Then the two terms
in the denominator $\propto \frac{1+{6\over Re_0}}{\omega}$ can be
neglected, and the extrema of $N(\omega)$ can be approximated by the extrema
of $\cos{\omega\tau}$,
\begin{equation}
\label{omr}
\omega_{r}(n)\simeq n{\pi\over\tau},\qquad n=0,\pm1,\pm2,...\ .
\end{equation} 
Now the amplitude of $\Delta$ is at maximum for frequencies
$\omega_{r}(n)$ with even $n$, and at minimum for $\omega_{r}(n)$ with
odd $n$. The distance between two maximum (or minimum)
amplitudes is $2\pi/\tau$ as indicated by the horizontal arrow in
the inset of Fig.\ref{logA}. For the first maxima and minima at
moderate frequencies this estimate is an approximation only; also 
their distances are not yet constant as they are for high
frequencies. 

\begin{figure}
\includegraphics[width=\columnwidth]{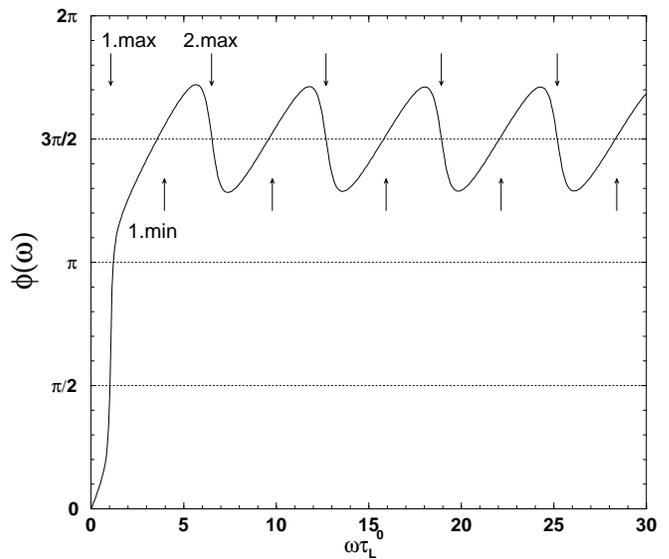}
\caption{ (a) Phase shift $\phi(\omega)$ 
as a function of the modulation frequency $\omega$
for weak modulation strength $e=0.1$, and $Re_0=10^4$. The upper
(lower) arrows indicate the frequencies of maximum (minimum) amplitude
of the response. For small $\omega$ the phase
$\phi(\omega)\propto \omega\tau$ behaves linearly.}
\label{Adel}
\end{figure}

In the high frequency limit, the oscillation of the response
at the frequencies $\omega_{r}$ of maximum or minimum amplitude is
phase shifted by $\phi_r(m)=(2m+1)\pi/2$, $m=\pm1,\pm3,...$:
\begin{equation}
\Delta(t)=e
\frac{({2\over3}+{6\over Re_0})}{\omega_r}\frac{(-1\pm
\alpha)\cos{\omega_r t}}{(1\mp\alpha)^2}\propto\sin{(\omega_r
t+\phi_r)}. 
\end{equation}
The prefactor $(-1\pm\alpha)$ is always negative, i.e., at the
response extrema we have $\Delta(t)\propto-\cos\omega_r
t=\sin(\omega_r t+\phi_r)$.  
In Fig.\ref{Adel} the phase shift $\phi(\omega)$, calculated from the
solution (\ref{dRt}), is shown as a function of the driving frequency
$\omega$ for all frequencies. As the phase shift starts with
$\phi(\omega=0)=0$ and changes continuously with increasing frequency,
we find that only $m=1$ is possible for the phase shift $\phi_r$ at
the response extrema. 
 The frequencies of the maximum and
minimum amplitudes of $\Delta$ are indicated by arrows. 
The only exception is the first maximum, where the approximation
for $\omega_r$, Eq. (\ref{omr}) does not yet hold. There, the phase shift is
near to $\pi/2$, corresponding to $m=0$.  
Another phase shift in this model is the one between the response
$\Delta(t)$ and the energy dissipation rate $\epsilon(t)$. According
to Eq. (\ref{eps2}) the dissipation rate is phase shifted by
$-\omega\tau$ with respect to the response $\Delta(t)$, i.e., this shift is
linearly growing with increasing frequency $\omega$. At the response
maxima and minima 
the phase shift is $-\omega_r\tau\simeq-n\pi$. 

The physics behind these response extrema can be explained as follows: The
time delay $\tau$ can be regarded as the (average) time which the input
energy needs before it is dissipated at small scales. 
In the case of maximum amplitude of
the response the time delay $\tau$ is a multiple $jT$ of the period
$T=2\pi/\omega$ of the forcing, whereas for the frequencies of minimum
amplitude the delay $\tau$ has an additional $T/2$. Therefore, at the
extrema of the response, the energy dissipation 
rate and the response are either in phase (maxima) or anti-phased
(minima). In the latter case the oscillation of the response is
strongly reduced. If, on the other hand, the driving frequency is such
that the response and the dissipation rate are in phase, the transport
of energy through the system is very effective and leads to an
enhanced oscillation. At the response maxima as well as at the minima
the phase shift between energy input rate and response is
$\phi_r=3\pi/2$.  

\subsection{Quality of the approximation for the delay $\tau$}
\label{tauapprox}

\begin{figure}
\includegraphics[width=\columnwidth]{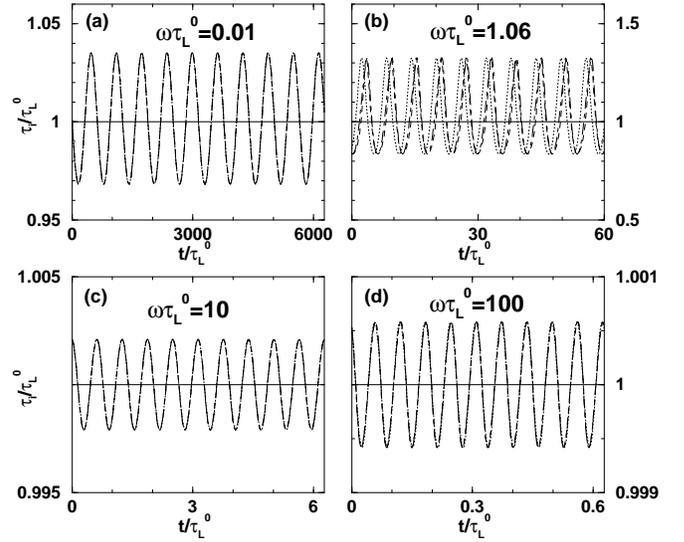}
\caption{ Successive approximations of the delay time $\tau$: First, constant 
approximation $\tau_0=a$ (solid lines); second, time dependent
approximation $\tau_1$ (dotted lines); third approximation $\tau_2$
(dashed lines); fourth approximation $\tau_3$ (dashed dotted lines)
for the delay time $\tau$, see Eqs. (\ref{tau1},\ref{taui}). (a)
$\omega\tau_L^0=0.01$. (b) $\omega\tau_L^0=1.06$. (c)
$\omega\tau_L^0=10$. (d) $\omega\tau_L^0=10^2$. In (a),(c) and (d) the
time dependent $\tau_i(t)$ for $i=1,2,3$ are indistinguishable.}
\label{tautest}
\end{figure}

In the above calculations we made an approximation for the time scale
$\tau$ of the cascade process. In the linearized model, we assumed
$\tau$ to be constant, $\tau=\tau_0=a$. Now we check a
posteriori the quality of this approximation. The solution (\ref{dRt})
of the linearized equation (\ref{lineq}) is used to compute the
``correct'' delay time $\tau$ step by step: The next approximation for
$\tau$ is 
\begin{equation}
\label{tau1}
\tau_1(t)=\frac{a}{\sqrt{1+\Delta(t)}},
\end{equation}
where the delay in Eq. (\ref{tauR}) is still neglected. Further steps
are:
\begin{eqnarray}
\label{taui}
\tau_2(t)&=&\frac{a}{\sqrt{1+\Delta(t-\tau_1)}},\\\nonumber
\tau_3(t)&=&\frac{a}{\sqrt{1+\Delta(t-\tau_2)}}, \quad\mbox{etc.}
\end{eqnarray} 
In Fig.\ref{tautest} $\tau_0=a$, $\tau_1$, $\tau_2$, and $\tau_3$ are
plotted for different frequencies. 
For $\omega=0.01$, the difference
between $\tau_1$, $\tau_2$, and $\tau_3$ is not visible. The variation
of the $\tau_i(t)$, ($i=1,2,3$), is largest at the frequency where the
amplitude of $\Delta$ is maximum, i.e., at $\omega\simeq {1\over
\tau_0}$. For all other frequencies, 
including at the response maxima, the variation of the $\tau_i(t)$ is 
much smaller than $\tau_0$ and $1/\omega$. At these frequencies 
it seems reasonable to approximate $\tau$ by the constant $\tau_0=a$. 
In Eq. (\ref{lineq}) the delay $\tau$ enters
into two terms, in $\propto \partial_t\Delta(t-\tau)$ on the lhs, and in 
$\propto\Delta(t-\tau)$ on the rhs. We calculate the relative error of these 
terms if $\tau=\tau_0$ instead of $\tau=\tau_i$ ($i=1,2,3$) is employed, 
using the solution (\ref{dRt}) for $\Delta$:
\begin{equation}
\label{error1}
\delta_1(\!\tau_i\!)\!=\!\sqrt{\!\frac{\int_0^{2\pi/\omega}\!
[\cos\big(\omega(t\!-\!\tau_0)\!+\!\phi\big)\!-\!
\cos\big(\omega(t\!-\!\tau_i)\!+\!\phi\big)]^2dt}
{\int_0^{2\pi/\omega}\!
\cos^2\big(\!\omega(t\!-\!\tau_0)\!+\!\phi\big)\ dt}}, 
\end{equation}
for the term on the lhs, and 
\begin{equation}
\label{error2}
\delta_2(\!\tau_i\!)\!=\!\sqrt{\!\frac{\int_0^{2\pi/\omega}\!
[\sin\big(\omega(t\!-\!\tau_0)\!+\!\phi\big)\!-\!
\sin\big(\omega(t\!-\!\tau_i)\!+\!\phi\big)]^2dt}
{\int_0^{2\pi/\omega}\!
\sin^2\big(\!\omega(t\!-\!\tau_0)\!+\!\phi\big)\ dt}}, 
\end{equation}
for the term on the rhs. 
The errors $\delta_1$ and $\delta_2$ are summarized in table
\ref{taberror} for the four chosen frequencies of Fig.\ref{tautest}. 
As expected, the errors are largest for the frequency with maximum response 
amplitude, $\omega=1.06$, where it becomes up to $23
\%$. For $\omega=10$ and beyond it is between $1$ and $5\%$.  
If one would allow for a time dependence of
$\tau$ in Eq. (\ref{lineq}) the response maxima would probably
become broader, possibly less pronounced. However,  within this mean
field theory we anyhow can make only approximate statements about the
frequencies and the values of the amplitudes at the response maxima. 
Namely, in 
the mean field approach the effects of the fluctuations on the
structure function are neglected. 
Therefore we believe that even with this approximation for the delay
time 
$\tau$ we can qualitatively predict the basic features of the system, 
which are the decrease of the amplitude of the response for high modulation 
frequencies, and the existence of response maxima at certain
frequencies due to the finite  
time needed by the energy cascade process.
\begin{table}
\begin{ruledtabular}
\begin{tabular}{|l|c|c|c|}
 & $\delta_1(\tau_1)$ & $\delta_1(\tau_2)$ &$\delta_1(\tau_3)$
\\\hline
$\omega\tau_L^0=0.01$ & $2.9\times10^{-4}$ &$2.9\times10^{-4}$ &$2.9\times10^{-4}$  
\\\hline
$\omega\tau_L^0=1.06$ &0.15 & 0.22  &0.23 
\\\hline
$\omega\tau_L^0=10$ &$0.016$ &$0.016$ & $0.016$
\\\hline
$\omega\tau_L^0=100$ & $0.046$&$0.046$ &$0.046$ 
\\
\hline
\hline
 & $\delta_2(\tau_1)$ & $\delta_2(\tau_2)$ &$\delta_2(\tau_3)$
\\\hline
$\omega\tau_L^0=0.01$ &$1.7\times10^{-4}$ &$1.7\times10^{-4}$ &$1.7\times10^{-4}$
\\\hline
$\omega\tau_L^0=1.06$ &0.22 &0.11 &0.12 
\\\hline
$\omega\tau_L^0=10$ &0.013  &0.013  &0.013 
\\\hline
$\omega\tau_L^0=100$ & $0.036$&$0.036$ &$0.036$ 
\\
\end{tabular}
\end{ruledtabular}
\caption{Relative errors $\delta_1$, $\delta_2$ according to Eqs.
(\ref{error1}),(\ref{error2}) made in the two relevant terms of Eq. 
(\ref{lineq}) by using the constant time delay $\tau_0=a$ instead of the
higher order approximations $\tau_i(t)$ for $\tau$.}
\label{taberror}
\end{table}
The validity of the approximation 
for $\tau$ will improve for smaller amplitudes $e$ of the modulation. However, 
for smaller $e$ the total amplitude $eA$ of the response will decrease as 
well and finally the amplitude of the response maxima and minima will
become so small that, in experiments or numerical simulations, 
the fluctuations will be larger than the maxima and minima. 

\section{An alternative argument to introduce the time-delay}
\label{EGleichung}

The energy balance equation (\ref{effgngl1}) and the expression for
the energy dissipation rate $\epsilon$, Eq. (\ref{eps1}), hold
for stationary systems. The time dependence of the quantities in these
equations in section \ref{method} has been introduced a posteriori by
arguments based on the picture of the energy cascade. It was not derived 
from the Navier Stokes equation, but is a
modeling ansatz. Therefore, there are several arguments to introduce this
time dependence. We want to discuss here another way of arguments 
which leads to a
slightly different equation for the response. 
The idea is to start from an equation which is already integrated
over all scales, i.e., does not depend on the scale $r$ any more in 
contrast to Eq. (\ref{effgngl1}) in section \ref{method}. 
The total energy per unit mass of the flow is $E\simeq 3u_{1,rms}^2/2$. It
is basically determined by the energy of the large scales. The
change with time of this energy equals the dissipation rate and the
energy input rate:
\begin{equation}
\label{E}
{d\over dt}E(t)=-\epsilon(t)+e_{in}(t).
\end{equation}
As the energy needs a time $\tau$ to travel down the eddy cascade
before it is dissipated, $\epsilon$ at time $t$ may be expressed
with Eq. (\ref{eps2}) and $E={1\over 4}D(L)$ by the total energy
$E$ at time $t-\tau$:
\begin{displaymath}
\epsilon(t)= c_{\epsilon}(E(t-\tau))\left({2\over
3}\right)^{3/2}\frac{E(t-\tau)^{3/2}}{L}. 
\end{displaymath} 
Together with the approximation for $c_{\epsilon}$, Eq. (\ref{ceps}), we get: 
\begin{equation}
\label{EGl}
{d\over dt}E(t)=-\frac{(4E(t-\tau))^{3/2}}{b^{3/2}L}
-6\frac{\nu}{L^2}E(t-\tau)+e_{in}(t). 
\end{equation}
As in section \ref{method} we express the energy $E$ by the Reynolds
number, 
$E=\frac{3\nu^2}{2L^2}Re^2$, write the energy input in terms of the
stationary Reynolds number $Re_0$, Eq. (\ref{re0gl}), and introduce
the reduced Reynolds number, $R(t)=Re(t)/Re_0$. Then, in time units of
$\tau_L^0$:
\begin{eqnarray}
\frac{d}{dt}R^2(t) 
&=&-{2\over 3}\left({6\over b}\right)^{3/2}\!(R^2(t-\tau))^{3/2}
-{6\over Re_0}R^2(t-\tau)
\nonumber\\\label{regl3}
&&+\left({2\over 3}\left({6\over
b}\right)^{3/2}+{6\over Re_0}\right)(1+e\sin{\omega t}). 
\end{eqnarray}
The only difference between this equation and the previous one, derived in
section \ref{method} (Eq. (\ref{regl2})), is that here the term $\propto
dR^2(t-\tau)/dt$ is missing. 

\begin{figure}
\includegraphics[width=\columnwidth]{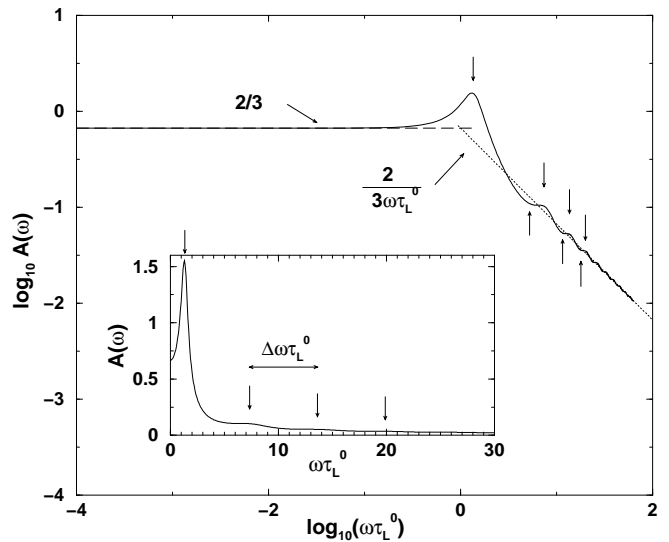}
\caption{Amplitude $A$ of the response $\Delta(t)$ as a
function of the driving frequency $\omega$ (log-log-scale) for weak
modulations ($e=0.1$) of
the input rate $e_{in}$, and $Re_0=10^4$ calculated from Eq. (\ref{regl3})
in linear approximation. The dashed line denotes the low
frequency limit of the oscillation amplitude, 2/3,  and the
dotted line corresponds to the mean trend of the high frequency limit,
$\frac{2}{3\omega\tau_L^0}$.
Inset: linear-scale-plot of
the response amplitude versus frequency. The small arrows indicate
the frequencies of maximum amplitude $\omega_r$ (in units of $\tau_L^0$)
calculated from the minima of the denominator in 
Eq. (\ref{dRt2}). The horizontal arrow denotes the
frequency distance $\Delta\omega$ (in units of $\tau_L^0$) between two frequencies for which the
amplitude is maximal (or minimal). It is $\Delta\omega\simeq
2\pi/\tau$ for high frequencies.}
\label{logAE} 
\end{figure}

If we solve Eq. (\ref{regl3}) within the same linear
approximation as employed in section \ref{response} for Eq. (\ref{regl2}),  we find the same features for the response,
see Fig. \ref{logAE}. 
The solution of the linearized equation obtained from Eq. (\ref{regl3}) 
reads:
\begin{equation}
\label{dRt2}
\Delta(t)=e
\frac{({2\over 3}+{6\over Re_0})}{\omega}
\frac{\Big[-\cos{\omega
t}+\frac{1+{6\over Re_0}}{\omega}\sin{\Big(\omega(t+\tau)\Big)}\Big]}
{1+(\frac{1+{6\over Re_0}}{\omega})^2-2\frac{1+{6\over
Re_0}}{\omega}\sin{\omega\tau}}. 
\end{equation}
Here, we have again set $b=6$ for simplicity. 
The response maxima are also observed, but they are less pronounced and
slightly shifted. The amplitude at the first (and strongest) maximum has 
only a value of $A_E\simeq 1.6$. In the linear response 
solution (\ref{dRt}) of the previous
model the terms originating from the second term on the lhs of
Eq. (\ref{lineq}) were responsible for the strong variation of the
amplitude at high frequencies. These terms are missing in the
present model. Therefore, we observe weaker amplitude maxima and minima
at high frequencies in this model, cf. Fig.\ref{logAE}.  If we take
the extended cascade time $\tau=a>1$ into acount,  e.g. $a=2.7$,
the response maxima are shifted towards smaller frequencies as
discussed in section \ref{resonances}. However, in this model, the heights
of all maxima including the first one is then slightly increased. At the
response maxima the energy cascade time scale $\tau$ and the
period of the driving modulation are not multiples of each other as
they are in the previous model, i.e., the response and the energy
dissipation rate are not exactly in phase. 
If one would observe the
response maxima in experiments or numerical simulations, one could
distinguish between the two models by studying the ratio between the
frequencies of the response maxima and the cascade time  scale $\tau$.
The phase shift $\phi$ between the energy input rate and the response 
becomes negative and
oscillates around $-\pi/2$ for higher frequencies. At the response
extrema 
it is near to $\phi_r\simeq -\pi/2$. Note that in the previous model
the phase shift was always positive. 

The two arguments to introduce the time delay are similar and are 
based on the same physical idea of a finite time lapse of the cascade 
process. However, we tend to prefer the first one, section
\ref{method}, because it introduces the time dependence at an earlier
stage. Eq. (\ref{effgngl1}) still resolves the scales
$r$ and it is therefore closer to the Navier Stokes equation than 
Eq. (\ref{E}). 

\section{Response of turbulent flow to a modulated driving force}
\label{modf}
In the previous sections we have studied the effect of a modulated energy 
input rate on turbulent flow. However, the energy input rate may not be 
a quantity which can be easily controlled in experiments. In some experiments 
it is more convenient to modulate the driving force instead. Then the 
resulting energy input rate as well as the total energy of the system 
can be considered as a response of the system. Therefore, in this section, 
we show how to treat this slightly modified case within the variable 
range mean-field theory and what differences we expect in these 
two different response functions. 

The derivation of Eq. (\ref{inteq}) for the response of the system in terms 
of the structure function $D(L,t)$ remains the same as explained in section 
\ref{method}. The energy input rate $e_{in}(t)$ in that equation is given by 
$e_{in}(t)=\langle\!\langle u_i^{(r)}(\x,t)f_i^{(r)}(\x,t)\rangle\!\rangle$. 
To introduce a modulated forcing instead of a modulated energy input rate, 
we therefore assume:
\ba
\label{einclosure}
e_{in}(t)&\simeq& D(L,t)^{1/2}f(t)
\\\nonumber
&=&D(L,t)^{1/2}f_0(1+e_f\sin\omega t)
\ea
Here, $f_0$ is the strength of the (stationary) forcing and 
$e_f$ the amplitude 
of the modulation. As has been discussed in section \ref{method}, we express 
the response in terms of the Reynolds number $Re(t)$ and relate the stationary 
Reynolds number $Re_0$ with the stationary forcing strength $f_0$, similar to 
Eq.  (\ref{re0gl}). Then we introduce the reduced Reynolds number 
$R(t)=\frac{Re(t)}{Re_0}$ and the dimensionless time $\tilde t=t/\tau_L^0$. The tilde  is dropped in the following.  The 
analogous equation to (\ref{regl2}) becomes: 
\begin{eqnarray}
\label{regl_modf}
\frac{d(R^2(t)\!-\!\alpha R^2(t\!-\!\tau))}{dt}\! 
&\!=\!&\!-{2\over 3}(R^2(t\!-\!\tau))
^{3/2}\\\nonumber
&&-\frac{6}{Re_0}R^2(t-\tau)
\\\nonumber
+&&(R^2(t))^{1/2}\!\left(\!{2\over 3}\!+\!\frac{6}{Re_0}\!\right)
\!(\!1\!+\!e_f\sin{\omega t}\!), 
\end{eqnarray}
where $b$ is set to $b=6$. 
We again assume small modulation amplitudes, i.e., $e_f\ll1$, and linearize 
Eq. (\ref{regl_modf}) in $\Delta(t)\equiv R^2(t)-1$. As before, the time delay 
$\tau$ is approximated by the time-independent constant $a$. With the same 
ansatz Eq.  (\ref{ansatz}) as in section \ref{trend} for 
modulated energy input rate, the linearized equation can be solved analytically, 
and the solution reads:
\begin{widetext}
\begin{equation}
\label{dRt_modf}
\Delta(t)=e_f
\frac{({2\over 3}+\frac{6}{Re_0})}{\omega}
\frac{\Big[-\cos{\omega t}+\alpha\cos{\Big(\omega
(t+\tau)\Big)}+\frac{1+{6\over Re_0}}{\omega}\sin{\Big(\omega(t+\tau)\Big)}
-\frac{({2\over 3}+\frac{6}{Re_0})}{2\omega}\sin{\omega t}\Big]}
{\Big[1+\alpha^2+(\frac{1+\frac{6}{Re_0}}{\omega})^2
+(\frac{({2\over 3}+\frac{6}{Re_0})}{2\omega})^2
-2\alpha\cos{\omega\tau}
-2\frac{1+\frac{6}{Re_0}}{\omega}\sin{\omega\tau}
+\frac{({2\over 3}+\frac{6}{Re_0})}{\omega}
(\alpha\sin\omega\tau-\frac{1+\frac{6}{Re_0}}{\omega}\cos{\omega\tau})\Big]}.
\end{equation}
\end{widetext}
This solution is very similar to the solution (\ref{dRt}) for a modulated 
energy input rate, but it contains some additional terms in both the numerator 
and the denominator. These terms only slightly modify the frequency dependence 
of the response $\Delta(t)$. 
\begin{figure}
\includegraphics[width=\columnwidth]{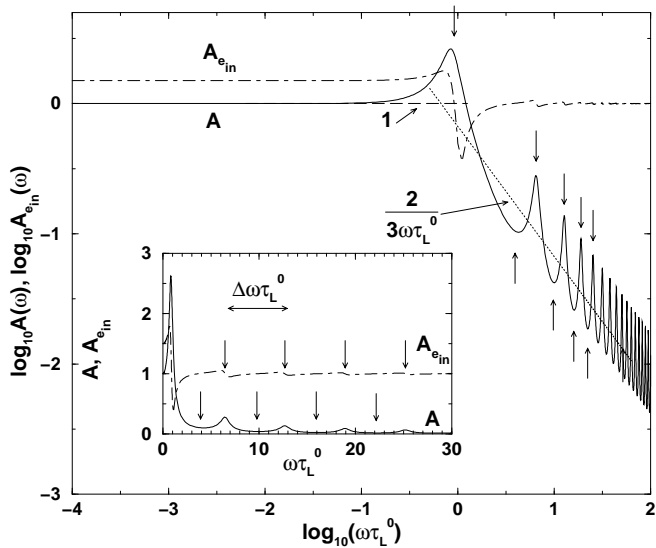}
\caption{Amplitude $A$ of the response $\Delta(t)$ as a
function of 
the driving frequency $\omega$ (log-log-scale) for weak modulations
($e_f=0.1$) of  the 
driving force $f$, and $Re_0=10^4$. The time scale of the energy cascade is set to $\tau=a=1$. The dashed line denotes the low
frequency limit of the oscillation amplitude, $1$, and the
dotted line corresponds to the mean trend of the high frequency limit, 
$\frac{2}{3}\frac{1}{\omega\tau_L^0}$. The dashed-dotted line represents the 
amplitude of the resulting energy input rate $e_{in}(t)$.  
Inset: linear-scale-plot of
the response amplitude (solid line) and the energy input amplitude 
(dashed-dotted line) versus frequency.  The small arrows indicate
the frequencies $\omega_r$ (in units of $\tau_L^0$) of the response
extrema 
calculated from the extrema of
the denominator in Eq. (\ref{dRt_modf}). The horizontal arrow denotes the
frequency distance $\Delta\omega$ (in units of $\tau_L^0$)  between two frequencies for which the
amplitude is maximal (or minimal). It is $\Delta\omega\simeq
2\pi/\tau$ for high frequencies. }
\label{logA_modf}
\end{figure}
In Fig.\ref{logA_modf} the amplitude $A(\omega)$ of the response $\Delta$ 
is plotted as a function of driving frequency for $Re_0=10^4$. As for the 
modulated energy input rate we note that the amplitude remains constant for 
low frequencies and decreases as $\propto 1/\omega$ for high frequencies. 
Also the response maxima and minima can be observed. Quantitatively, the 
low frequency limit for a modulated forcing is different from the modulated 
energy input rate case. For low driving frequencies, $\omega\ll 1/\tau$, we 
can approximate Eq. (\ref{dRt_modf}) by:
\be
\label{lowf_modf}
\Delta(t)\simeq e_f\frac{\frac{2}{3\omega}(\frac{1}{\omega}\sin{(\omega(t+\tau))}-\frac{1}{3\omega}\sin{\omega t})}{(\frac{1}{\omega})^2+(\frac{1}{3\omega})^2-2\frac{1}{\omega}\frac{1}{3\omega}\cos\omega\tau}.
\ee
The terms $6/Re_0 \ll 1$ have been omitted here for simplicity. 
In the limit $\omega\tau\rightarrow 0$, with $\sin{\omega\tau}\rightarrow 0$ and $\cos{\omega\tau}\rightarrow 1$, the amplitude $A$ 
of the response (cf. Eq. (\ref{ansatz})) becomes equal to one instead of 
two thirds (cf. Eq. (\ref{lowf})) for  a modulated energy input rate. 
The frequencies of the reponse maxima 
and minima are determined by the extrema of the denominator of the solution 
(\ref{dRt_modf}). They are slightly shifted as compared to the case with 
modulated energy input rate (Eq. (\ref{dRt})). The amplitude at the first maximum is smaller than in the case with modulated energy input rate, namely $A_E\simeq 2.7$.  
However, in the limit of very 
high driving frequencies, $\omega\gg\frac{1}{\tau}$, Eq.  (\ref{dRt_modf}) 
can be approximated by Eq. (\ref{env}), i.e., the response amplitudes of both 
cases become identical. 

It was pointed out in the beginning of this section that, if we modulate the 
driving force, the energy input rate is not a controlled quantity, but can be 
considered as well as a response of the system. This has been measured in a 
recent experimental study by Cadot et al.\cite{cad02}.  Within the mean-field 
theory the energy input rate for a modulated driving force can be calculated 
as: 
\be
\label{ein_modf}
\frac{e_{in}(t)}{e_{in,0}}=\sqrt{1+\Delta(t)}(1+e_f\sin\omega t),
\ee
where $e_{in,0}=\langle\!\langle D_{L,0}^{1/2}f_0\rangle\!\rangle$ is the 
stationary energy input rate for constant forcing without modulation. In order 
to extract the amplitude of the energy input rate, we fit it by a function of 
the form $\frac{e_{in}(t)}{e_{in,0}}=1+e_f 
A_{e_{in}}\sin(\omega(t+\phi))$. This 
is justified as long as the modulation amplitude $e_f$ is small, i.e., 
$e_f\ll 1$, because then $\Delta(t)$ is of the same order of magnitude as 
$e_f$ and Eq. (\ref{ein_modf}) can be approximated by 
$\frac{e_{in}(t)}{e_{in,0}}-1
\simeq{1\over 2}\Delta(t)+e_f\sin\omega t+O(\Delta^2)$. The amplitude 
$A_{e_{in}}$ of the energy input rate is included in Fig.\ref{logA_modf} 
as a dashed-dotted line. For low driving frequencies, $\omega\ll{1\over\tau}$, 
the amplitude $A_{e_{in}}$ is nearly constant and is  $3/2$, whereas for high 
frequencies it decreases and finally saturates at one. Also the response 
maxima can be observed in the energy input rate: At the same frequencies, 
where the response $\Delta$ shows amplitude maxima, we observe a maximum 
directly followed by a minimum in the amplitude of the energy input rate. 

In conclusion, if the driving force instead of the energy input rate is 
modulated, the general behavior of the response in terms of the second order 
structure function on the larges scale remains the same, 
including the response maxima and minima. For low driving frequencies, the 
amplitude of the response becomes equal to the amplitude of the forcing. In 
addition, the energy input rate can be regarded as a different measure for 
the response of the system, which also shows the response maxima at 
frequencies connected with the energy cascade time scale $\tau$. 

\section{Conclusions}
\label{discussion}
We calculated the response of isotropic and
homogeneous turbulence to a weak modulation of the energy input rate
$e_{in}$ within a mean-field theory. For
low frequencies the system follows the input rate modulation whereas for
high frequencies the amplitude of the response decreases $\propto
1/\omega$.  Due to the intrinsic time scale of the
system, the eddy-turnover time $\tau$, which also characterizes the energy
 transport time down the eddy cascade, there are certain frequencies,
$\omega_{r}\simeq n{\pi\over\tau}$, where the amplitude 
of the response is either  increased or decreased. At these
frequencies 
the phase shift $\phi$ between the energy input rate and the response
is $\phi_r\simeq 3\pi/2$. The response extrema occur
when the eddy-turnover time is an even or odd multiple of half
the modulation period $T/2=\pi/\omega$,  respectively. In the case
of response maxima, the energy dissipation rate
and the response of the system are in phase. This can be understood as
 a very effective transport of energy through the system.  At the
amplitude minima, 
instead, the response of the system is strongly reduced. Then, the
energy dissipation rate and the response are exactly anti-phased. 

In the mean-field approach the 
fluctuations of the energy flow rate through the system and of the
large eddy turnover time are
neglected. In experiment or numerical simulation the fluctuations are
however present. They may  lead
to broader and less pronounced response maxima, i.e., partly wash
out the response maxima and minima.

With increasing
modulation amplitude $e$ of the energy input rate the response maxima
are expected to become more significant due to the better signal to fluctuation ratio. But remember that for higher modulation
amplitudes $e$, 
the time scale of the eddy cascade, which enters into our model as a
time 
delay, becomes time dependent.  
This as well could lead to less pronounced
response maxima as discussed in section \ref{tauapprox}. 

A way to check if
the characteristic feature of the response maxima and minima can still
be well identified under the influence of fluctuations, would be to
perform numerical simulations of the Navier Stokes Equation with a
modulated driving. However, as not only high Reynolds numbers are
needed to achieve fully developed, isotropic and homogeneous
turbulence, but also the response as a function of time for a wide
range of frequencies has to be calculated, the computational effort
would be too high. Therefore, numerical simulations within two
dynamical cascade models of turbulence, the GOY shell model 
\cite{gle73,yam87,yam88a,ohk89,jen91,kad95,bif03}
and the reduced wave vector set approximation (REWA) 
\cite{egg91a,gnlo92b,gnlo94a}, were
performed \cite{hey03}. 
These models take into account the fluctuations. The basic
trend of the frequency dependence of the response amplitude as
calculated within the mean-field model can be reproduced in both
numerical models. We also clearly find the main maximum in both models although it os of course washed out by the fluctuations. The higher maxima and minima however seem to be completely washed out. 
 
Also a recent experimental study of modulated
turbulence by Cadot et al. \cite{cad02} showed evidence for the
existence of the response maxima. This experiment may be comparable with 
our study of a modulated driving force as discussed in section \ref{modf}. 
The response maxima were measured in the amplitude of the energy input
rate. In addition, a constant response amplitude for low driving
frequencies and a $1/\omega$-decay of the velocity response amplitude
for large frequencies has been observed. This is in agreement with the
$1/\omega$-decay of the energy response amplitude which we have found
in the mean-field model. The velocity response
$(u(t)-u_0)/u_0=\Delta_u(t)$, where $u(t)$ is the measured velocity
modulus and $u_0$ the (stationary) mean velocity, is connected to the
energy response $\Delta(t)$ which we have calculated in this paper
by $1+\Delta(t)=u(t)^2/u_0^2=(1+\Delta_u(t))^2\simeq
1+2\Delta_u(t)+O(\Delta_u^2)$. As only small modulation amplitudes are
considered the term $+O(\Delta_u^2)$ will be negligible because
$\Delta_u\ll1$. 
Therefore, the experimentally measured $1/\omega$-decay of 
the amplitude $\Delta_u$ of the velocity response 
is just what one would expect
from our theoretical prediction $\Delta\propto 1/\omega$ for the 
amplitude of the energy response.

We hope that the present work will stimulate even more experimental and numercial studies on
the role of the energy cascade time scale in modulated turbulence. 

\noindent
{\bf Acknowledgments:}
We thank R.\ Pandit and B.\ Eckhardt for very helpful discussions. 
The work is part of the research  program of the Stichting voor 
Fundamenteel Onderzoek der Materie (FOM), which is financially supported 
by the Nederlandse  Organisatie voor Wetenschappelijk Onderzoek (NWO).
This research was also supported 
by  the German-Israeli Foundation (GIF) 
and by the European Union under contract HPRN-CT-2000-00162.
\vspace{-0.5cm}


\begin{thebibliography}{32}
\expandafter\ifx\csname natexlab\endcsname\relax\def\natexlab#1{#1}\fi
\expandafter\ifx\csname bibnamefont\endcsname\relax
  \def\bibnamefont#1{#1}\fi
\expandafter\ifx\csname bibfnamefont\endcsname\relax
  \def\bibfnamefont#1{#1}\fi
\expandafter\ifx\csname citenamefont\endcsname\relax
  \def\citenamefont#1{#1}\fi
\expandafter\ifx\csname url\endcsname\relax
  \def\url#1{\texttt{#1}}\fi
\expandafter\ifx\csname urlprefix\endcsname\relax\def\urlprefix{URL }\fi
\providecommand{\bibinfo}[2]{#2}
\providecommand{\eprint}[2][]{\url{#2}}

\bibitem[{\citenamefont{Rippeth et~al.}(2001)\citenamefont{Rippeth, Fisher, and
  Simpson}}]{rip01}
\bibinfo{author}{\bibfnamefont{T.~P.} \bibnamefont{Rippeth}},
  \bibinfo{author}{\bibfnamefont{N.~R.} \bibnamefont{Fisher}},
  \bibnamefont{and} \bibinfo{author}{\bibfnamefont{J.~H.}
  \bibnamefont{Simpson}}, \bibinfo{journal}{J. Phys. Oceanogr.}
  \textbf{\bibinfo{volume}{31}}, \bibinfo{pages}{2458} (\bibinfo{year}{2001}).

\bibitem[{\citenamefont{Scotti and Piomelli}(2001)}]{sco01}
\bibinfo{author}{\bibfnamefont{A.}~\bibnamefont{Scotti}} \bibnamefont{and}
  \bibinfo{author}{\bibfnamefont{U.}~\bibnamefont{Piomelli}},
  \bibinfo{journal}{Phys. Fluids} \textbf{\bibinfo{volume}{13}},
  \bibinfo{pages}{1367} (\bibinfo{year}{2001}).

\bibitem[{\citenamefont{Qiu et~al.}(2000)\citenamefont{Qiu, Yao, and
  Tong}}]{qiu00a}
\bibinfo{author}{\bibfnamefont{X.~L.} \bibnamefont{Qiu}},
  \bibinfo{author}{\bibfnamefont{S.~H.} \bibnamefont{Yao}}, \bibnamefont{and}
  \bibinfo{author}{\bibfnamefont{P.}~\bibnamefont{Tong}},
  \bibinfo{journal}{Phys. Rev. E} \textbf{\bibinfo{volume}{61}},
  \bibinfo{pages}{R6075} (\bibinfo{year}{2000}).

\bibitem[{\citenamefont{Qiu and Tong}(2001{\natexlab{a}})}]{qiu01b}
\bibinfo{author}{\bibfnamefont{X.~L.} \bibnamefont{Qiu}} \bibnamefont{and}
  \bibinfo{author}{\bibfnamefont{P.}~\bibnamefont{Tong}},
  \bibinfo{journal}{Phys. Rev. E} \textbf{\bibinfo{volume}{64}},
  \bibinfo{pages}{036304} (\bibinfo{year}{2001}{\natexlab{a}}).

\bibitem[{\citenamefont{Qiu and Tong}(2001{\natexlab{b}})}]{qiu01a}
\bibinfo{author}{\bibfnamefont{X.~L.} \bibnamefont{Qiu}} \bibnamefont{and}
  \bibinfo{author}{\bibfnamefont{P.}~\bibnamefont{Tong}},
  \bibinfo{journal}{Phys. Rev. Lett} \textbf{\bibinfo{volume}{87}},
  \bibinfo{pages}{094501} (\bibinfo{year}{2001}{\natexlab{b}}).

\bibitem[{\citenamefont{Labb\'e et~al.}(1996)\citenamefont{Labb\'e, Pinton, and
  Fauve}}]{lab96b}
\bibinfo{author}{\bibfnamefont{R.}~\bibnamefont{Labb\'e}},
  \bibinfo{author}{\bibfnamefont{J.~F.} \bibnamefont{Pinton}},
  \bibnamefont{and} \bibinfo{author}{\bibfnamefont{S.}~\bibnamefont{Fauve}},
  \bibinfo{journal}{Phys. Fluids} \textbf{\bibinfo{volume}{8}},
  \bibinfo{pages}{914} (\bibinfo{year}{1996}).

\bibitem[{\citenamefont{Auma\^{\i}tre et~al.}(2000)\citenamefont{Auma\^{\i}tre,
  Fauve, and Pinton}}]{aum00}
\bibinfo{author}{\bibfnamefont{S.}~\bibnamefont{Auma\^{\i}tre}},
  \bibinfo{author}{\bibfnamefont{S.}~\bibnamefont{Fauve}}, \bibnamefont{and}
  \bibinfo{author}{\bibfnamefont{J.~F.} \bibnamefont{Pinton}},
  \bibinfo{journal}{Eur. Phys. J. B} \textbf{\bibinfo{volume}{16}},
  \bibinfo{pages}{563} (\bibinfo{year}{2000}).

\bibitem[{\citenamefont{Lohse}(2000)}]{loh00}
\bibinfo{author}{\bibfnamefont{D.}~\bibnamefont{Lohse}},
  \bibinfo{journal}{Phys. Rev. E} \textbf{\bibinfo{volume}{62}},
  \bibinfo{pages}{4946} (\bibinfo{year}{2000}).

\bibitem[{\citenamefont{Hooghoudt et~al.}(2001)\citenamefont{Hooghoudt, Lohse,
  and Toschi}}]{hog01}
\bibinfo{author}{\bibfnamefont{J.~O.} \bibnamefont{Hooghoudt}},
  \bibinfo{author}{\bibfnamefont{D.}~\bibnamefont{Lohse}}, \bibnamefont{and}
  \bibinfo{author}{\bibfnamefont{F.}~\bibnamefont{Toschi}},
  \bibinfo{journal}{Phys. Fluids} \textbf{\bibinfo{volume}{13}},
  \bibinfo{pages}{2013} (\bibinfo{year}{2001}).

\bibitem[{\citenamefont{Effinger and Grossmann}(1987)}]{eff87}
\bibinfo{author}{\bibfnamefont{H.}~\bibnamefont{Effinger}} \bibnamefont{and}
  \bibinfo{author}{\bibfnamefont{S.}~\bibnamefont{Grossmann}},
  \bibinfo{journal}{Z. Phys. B} \textbf{\bibinfo{volume}{66}},
  \bibinfo{pages}{289} (\bibinfo{year}{1987}).

\bibitem[{\citenamefont{Monin and Yaglom}(1975)}]{my75}
\bibinfo{author}{\bibfnamefont{A.~S.} \bibnamefont{Monin}} \bibnamefont{and}
  \bibinfo{author}{\bibfnamefont{A.~M.} \bibnamefont{Yaglom}},
  \emph{\bibinfo{title}{Statistical Fluid Mechanics}} (\bibinfo{publisher}{The
  MIT Press}, \bibinfo{address}{Cambridge, Massachusetts},
  \bibinfo{year}{1975}).

\bibitem[{\citenamefont{Sreenivasan}(1995)}]{sre95}
\bibinfo{author}{\bibfnamefont{K.~R.} \bibnamefont{Sreenivasan}},
  \bibinfo{journal}{Phys. Fluids} \textbf{\bibinfo{volume}{7}},
  \bibinfo{pages}{2778} (\bibinfo{year}{1995}).

\bibitem[{\citenamefont{Sreenivasan and Antonia}(1997)}]{sre97}
\bibinfo{author}{\bibfnamefont{K.~R.} \bibnamefont{Sreenivasan}}
  \bibnamefont{and} \bibinfo{author}{\bibfnamefont{R.~A.}
  \bibnamefont{Antonia}}, \bibinfo{journal}{Ann. Rev. of Fluid Mech.}
  \textbf{\bibinfo{volume}{29}}, \bibinfo{pages}{435} (\bibinfo{year}{1997}).

\bibitem[{\citenamefont{Pope}(2000)}]{pop00}
\bibinfo{author}{\bibfnamefont{S.~B.} \bibnamefont{Pope}},
  \emph{\bibinfo{title}{Turbulent Flows}} (\bibinfo{publisher}{Cambridge
  University Press}, \bibinfo{address}{Cambridge}, \bibinfo{year}{2000}).

\bibitem[{\citenamefont{Sreenivasan}(1984)}]{sre84}
\bibinfo{author}{\bibfnamefont{K.~R.} \bibnamefont{Sreenivasan}},
  \bibinfo{journal}{Phys. Fluids} \textbf{\bibinfo{volume}{27}},
  \bibinfo{pages}{1048} (\bibinfo{year}{1984}).

\bibitem[{\citenamefont{Sreenivasan}(1998)}]{sre98}
\bibinfo{author}{\bibfnamefont{K.~R.} \bibnamefont{Sreenivasan}},
  \bibinfo{journal}{Phys. Fluids} \textbf{\bibinfo{volume}{10}},
  \bibinfo{pages}{528} (\bibinfo{year}{1998}).

\bibitem[{\citenamefont{Lohse}(1994)}]{loh94a}
\bibinfo{author}{\bibfnamefont{D.}~\bibnamefont{Lohse}},
  \bibinfo{journal}{Phys. Rev. Lett.} \textbf{\bibinfo{volume}{73}},
  \bibinfo{pages}{3223} (\bibinfo{year}{1994}).

\bibitem[{\citenamefont{Grossmann}(1995)}]{gro95}
\bibinfo{author}{\bibfnamefont{S.}~\bibnamefont{Grossmann}},
  \bibinfo{journal}{Phys. Rev. E} \textbf{\bibinfo{volume}{51}},
  \bibinfo{pages}{6275} (\bibinfo{year}{1995}).

\bibitem[{\citenamefont{Stolovitzky and Sreenivasan}(1995)}]{sre95d}
\bibinfo{author}{\bibfnamefont{G.}~\bibnamefont{Stolovitzky}} \bibnamefont{and}
  \bibinfo{author}{\bibfnamefont{K.~R.} \bibnamefont{Sreenivasan}},
  \bibinfo{journal}{Phys. Rev. E} \textbf{\bibinfo{volume}{52}},
  \bibinfo{pages}{3242} (\bibinfo{year}{1995}).

\bibitem[{\citenamefont{Rao et~al.}(1990)\citenamefont{Rao, Krishnamurthy, and
  Pandit}}]{rao90}
\bibinfo{author}{\bibfnamefont{M.}~\bibnamefont{Rao}},
  \bibinfo{author}{\bibfnamefont{H.}~\bibnamefont{Krishnamurthy}},
  \bibnamefont{and} \bibinfo{author}{\bibfnamefont{R.}~\bibnamefont{Pandit}},
  \bibinfo{journal}{Phys. Rev. B} \textbf{\bibinfo{volume}{42}},
  \bibinfo{pages}{856} (\bibinfo{year}{1990}).

\bibitem[{\citenamefont{Cadot et~al.}(2002)\citenamefont{Cadot, Titon, and
  Bonn}}]{cad02}
\bibinfo{author}{\bibfnamefont{O.}~\bibnamefont{Cadot}},
  \bibinfo{author}{\bibfnamefont{J.~H.} \bibnamefont{Titon}}, \bibnamefont{and}
  \bibinfo{author}{\bibfnamefont{D.}~\bibnamefont{Bonn}},
  \bibinfo{journal}{Preprint, submitted to J. Fluid Mech.}
  (\bibinfo{year}{2002}).

\bibitem[{\citenamefont{Gledzer}(1973)}]{gle73}
\bibinfo{author}{\bibfnamefont{E.~B.} \bibnamefont{Gledzer}},
  \bibinfo{journal}{Sov. Phys. Dokl.} \textbf{\bibinfo{volume}{18}},
  \bibinfo{pages}{216} (\bibinfo{year}{1973}).

\bibitem[{\citenamefont{Yamada and Ohkitani}(1987)}]{yam87}
\bibinfo{author}{\bibfnamefont{M.}~\bibnamefont{Yamada}} \bibnamefont{and}
  \bibinfo{author}{\bibfnamefont{K.}~\bibnamefont{Ohkitani}},
  \bibinfo{journal}{J. Phys. Soc. Jpn.} \textbf{\bibinfo{volume}{56}},
  \bibinfo{pages}{4210} (\bibinfo{year}{1987}).

\bibitem[{\citenamefont{Yamada and Ohkitani}(1988)}]{yam88a}
\bibinfo{author}{\bibfnamefont{M.}~\bibnamefont{Yamada}} \bibnamefont{and}
  \bibinfo{author}{\bibfnamefont{K.}~\bibnamefont{Ohkitani}},
  \bibinfo{journal}{Prog. Theor. Phys.} \textbf{\bibinfo{volume}{79}},
  \bibinfo{pages}{1265} (\bibinfo{year}{1988}).

\bibitem[{\citenamefont{Ohkitani and Yamada}(1989)}]{ohk89}
\bibinfo{author}{\bibfnamefont{K.}~\bibnamefont{Ohkitani}} \bibnamefont{and}
  \bibinfo{author}{\bibfnamefont{M.}~\bibnamefont{Yamada}},
  \bibinfo{journal}{Prog. Theor. Phys.} \textbf{\bibinfo{volume}{81}},
  \bibinfo{pages}{329} (\bibinfo{year}{1989}).

\bibitem[{\citenamefont{Kadanoff et~al.}(1995)\citenamefont{Kadanoff, Lohse,
  Wang, and Benzi}}]{kad95}
\bibinfo{author}{\bibfnamefont{L.}~\bibnamefont{Kadanoff}},
  \bibinfo{author}{\bibfnamefont{D.}~\bibnamefont{Lohse}},
  \bibinfo{author}{\bibfnamefont{J.}~\bibnamefont{Wang}}, \bibnamefont{and}
  \bibinfo{author}{\bibfnamefont{R.}~\bibnamefont{Benzi}},
  \bibinfo{journal}{Phys. Fluids} \textbf{\bibinfo{volume}{7}},
  \bibinfo{pages}{617} (\bibinfo{year}{1995}).

\bibitem[{\citenamefont{Biferale}(2003)}]{bif03}
\bibinfo{author}{\bibfnamefont{L.}~\bibnamefont{Biferale}},
  \bibinfo{journal}{Ann.~Rev. Fluid Mech.} \textbf{\bibinfo{volume}{35}},
  \bibinfo{pages}{441} (\bibinfo{year}{2003}).

\bibitem[{\citenamefont{Jensen et~al.}(1991)\citenamefont{Jensen, Paladin, and
  Vulpiani}}]{jen91}
\bibinfo{author}{\bibfnamefont{M.~H.} \bibnamefont{Jensen}},
  \bibinfo{author}{\bibfnamefont{G.}~\bibnamefont{Paladin}}, \bibnamefont{and}
  \bibinfo{author}{\bibfnamefont{A.}~\bibnamefont{Vulpiani}},
  \bibinfo{journal}{Phys. Rev. A} \textbf{\bibinfo{volume}{43}},
  \bibinfo{pages}{798} (\bibinfo{year}{1991}).

\bibitem[{\citenamefont{Grossmann and Lohse}(1994)}]{gnlo94a}
\bibinfo{author}{\bibfnamefont{S.}~\bibnamefont{Grossmann}} \bibnamefont{and}
  \bibinfo{author}{\bibfnamefont{D.}~\bibnamefont{Lohse}},
  \bibinfo{journal}{Phys. Fluids} \textbf{\bibinfo{volume}{6}},
  \bibinfo{pages}{611} (\bibinfo{year}{1994}).

\bibitem[{\citenamefont{Eggers and Grossmann}(1991)}]{egg91a}
\bibinfo{author}{\bibfnamefont{J.}~\bibnamefont{Eggers}} \bibnamefont{and}
  \bibinfo{author}{\bibfnamefont{S.}~\bibnamefont{Grossmann}},
  \bibinfo{journal}{Phys. Fluids A} \textbf{\bibinfo{volume}{3}},
  \bibinfo{pages}{1958} (\bibinfo{year}{1991}).

\bibitem[{\citenamefont{Grossmann and Lohse}(1992)}]{gnlo92b}
\bibinfo{author}{\bibfnamefont{S.}~\bibnamefont{Grossmann}} \bibnamefont{and}
  \bibinfo{author}{\bibfnamefont{D.}~\bibnamefont{Lohse}}, \bibinfo{journal}{Z.
  Phys. B} \textbf{\bibinfo{volume}{89}}, \bibinfo{pages}{11}
  (\bibinfo{year}{1992}).

\bibitem[{\citenamefont{von~der Heydt et~al.}(2003)\citenamefont{von~der Heydt,
  Grossmann, and Lohse}}]{hey03}
\bibinfo{author}{\bibfnamefont{A.}~\bibnamefont{von~der Heydt}},
  \bibinfo{author}{\bibfnamefont{S.}~\bibnamefont{Grossmann}},
  \bibnamefont{and} \bibinfo{author}{\bibfnamefont{D.}~\bibnamefont{Lohse}},
  \bibinfo{journal}{Preprint, submitted to Phys. Rev. E}  (\bibinfo{year}{2003}).

\end{thebibliography}

\end{document}